\documentclass[twocolumn]{aastex61}
\usepackage{amsmath}

\def\HII{{\ion{H}{2}}}
\def\OII{[{\ion{O}{2}}]}
\def\4959_5007{[\ion{O}{3}]~$\lambda \lambda$4959,5007}
\def\OIII49595007{[\ion{O}{3}]~$\lambda \lambda 4959,5007$}
\def\ratioR23{([\ion{O}{2}]~$\lambda \lambda$3727,9 + [\ion{O}{3}]~$\lambda\lambda$4959,5007)/H$\beta$}
\def\R23{${\rm R}_{23}$}
\def\dS23{${\rm S}_{23}$}

\def\NII{[{\ion{N}{2}}]}
\def\OIIIOII{[\ion{O}{3}]/[\ion{O}{2}]}
\def\NIIOII{[\ion{N}{2}]/[\ion{O}{2}]}

\def\ratioS23{([\ion{S}{2}]~$\lambda \lambda$6717,31 +[\ion{S}{3}]~$\lambda\lambda$9069,9532)/H$\beta$}

\def\SII{[{\ion{S}{2}}]}

\def\Hb{{H$\beta$}}
\def\O4363{[{\ion{O}{3}}]~$\lambda$4363}
\def\OIII{[{\ion{O}{3}}]}

\def\Ha{{H$\alpha$}}
\def\Ep{$E_{\rm peak}$}
\def\fN{$f_{\rm NLR}$}

\accepted{February 16, 2019 for publication in ApJ}

\shorttitle{AGN metallicity in SDSS}
\shortauthors{A. D. Thomas et al.}

\begin{document}

\title{The mass-metallicity relation of local active galaxies}

\correspondingauthor{Adam D. Thomas}
\email{adam.thomas@anu.edu.au}

\author[0000-0003-1761-6533]{Adam D. Thomas}
\author{Lisa J. Kewley}
\author{Michael A. Dopita}
\author{Brent A. Groves}
\affiliation{RSAA, Australian National University, Cotter Road, Weston Creek, ACT 2611, Australia}
\affiliation{ARC Centre of Excellence for All Sky Astrophysics in 3 Dimensions (ASTRO 3D), Cotter Road, Weston Creek, ACT 2611, Australia}

\author{Andrew M. Hopkins}
\affiliation{Australian Astronomical Observatory, 105 Delhi Rd, North Ryde, NSW 2113, Australia}
\affiliation{ARC Centre of Excellence for All Sky Astrophysics in 3 Dimensions (ASTRO 3D), 105 Delhi Rd, North Ryde, NSW 2113, Australia}

\author{Ralph S. Sutherland}
\affiliation{RSAA, Australian National University, Cotter Road, Weston Creek, ACT 2611, Australia}
\affiliation{ARC Centre of Excellence for All Sky Astrophysics in 3 Dimensions (ASTRO 3D), Cotter Road, Weston Creek, ACT 2611, Australia}


 
\begin{abstract}
We systematically measure the gas-phase metallicities and the mass-metallicity relation of a large sample of local active galaxies for the first time.  Observed emission-line fluxes from the Sloan Digital Sky Survey (SDSS) are compared to a four-dimensional grid of photoionization models using the Bayesian parameter estimation code NebulaBayes.  For the first time we take into account arbitrary mixing between \HII\ region and narrow-line region (NLR) emission, and the models are also varied with metallicity, ionization parameter in the NLR, and the gas pressure.  The active galactic nucleus (AGN) oxygen abundance is found to increase by $\Delta {\rm O/H} \sim 0.1$~dex as a function of host galaxy stellar mass over the range $10.1 < \log M_* / M_\odot < 11.3$.  We also measure the metallicity and ionization parameter of 231000 star-forming galaxies for comparison with the sample of 7670 Seyfert~2 galaxies.  A systematic offset in oxygen abundance of 0.09~dex is observed between the mass-metallicity relations of the star-forming and active galaxies.  We investigate potential causes of the offset, including sample selection and the treatment in the models of diffuse ionized gas, pressure, and ionization parameter.  We cannot identify the major cause(s), but suspect contributions due to deficiencies in modeling the ionizing spectra and the treatment of dust physics.  Optical diagnostic diagrams are presented with the star-forming and Seyfert data colored by the inferred oxygen abundance, ionization parameter and gas pressure, clearly illustrating the trends in these quantities.
\end{abstract}


\section{Introduction} \label{sec:Intro}

The gas-phase metallicity of a galaxy is the result of enrichment by stellar processes and the exchange of material with its environment \citep[e.g.][]{Lilly_2013_bathtub}.  The assembled stellar mass of a galaxy is known to be strongly positively correlated with its metallicity \citep[e.g.][]{Lequeux_1979_MZ, Tremonti_2004_MZ}.

This mass-metallicity relation \citep[or, similarly, the luminosity-metallicity relation; e.g.][]{Zaritsky_1994_Luminosity_Z} has been measured for star-forming galaxies both locally \citep[e.g.][]{Tremonti_2004_MZ} and at a range of redshifts \citep[e.g.][]{Zahid_2014_MZ, Sanders_2015_MOSDEF_MZ, Wuyts_2016_highz_Z}.  For galaxies hosting active galactic nuclei (AGN), \citet{Matsuoka_2018_NLR_MZ} recently found a positive correlation between metallicity and host galaxy stellar mass at high redshift.  However, to date the mass-metallicity relation has not been cleanly measured for active galaxies in the local universe.

Metallicity measurements in AGN have typically focused on the broad-line regions (BLRs) of Seyfert AGN and quasars \citep{Hamann_and_Ferland_1999_QSO_Z_review}, although narrow-line region (NLR) abundances have also received some attention.  Inferred quasar abundances are ubiquitously high and show negligible evolution with redshift when considering either the BLRs \citep[e.g.][]{Dietrich_2003_BLR_Z, Nagao_2006_SDSS_BLR_Z, Juarez_2009_QSO_Z} or the NLRs \citep{Nagao_2006_AGN_NLR_Z, Dors_2014_AGN_Z}.  Quasar luminosity is correlated with the metallicity measured in both the BLR \citep{Nagao_2006_SDSS_BLR_Z} and in the NLR \citep{Nagao_2006_AGN_NLR_Z}.  There is a strong correlation between metallicity-sensitive NLR and BLR line ratios in the same objects  \citep{Du_2014_NLR_BLR_Z}.
	
Optical emission-line metallicity diagnostics for NLRs include those of \citet{Storchi-Bergmann_1998_AGN_Z} and \citet{Castro_2017_AGN_Z}.  Metallicity diagnostics for NLRs are reviewed by \citet{Dors_2015_AGN_Z}.  We note that independent estimates of nuclear metallicities in active galaxies may be obtained from the metallicities measured in nearby \HII\ regions and by radial extrapolation of \HII\ region metallicities from the galactic disk \cite[e.g.][]{Storchi-Bergmann_1998_AGN_Z, Dopita_2014_probing_ENLR_I_S7}.

The mixing of \HII\ region and NLR excitation in Seyfert spectra is a key complication in the use of metallicity diagnostics.  The analysis in this study compares observations to `mixed' models that account for varying contributions from \HII\ region and NLR emission.  Here, using a statistically powerful sample from the Sloan Digital Sky Survey \citep[SDSS;][]{York_2000_SDSS} and accounting for \HII-AGN mixing, we are able to investigate the AGN mass-metallicity relation in detail for the first time.

In Paper~1 (Thomas et al. 2018, accepted to ApJL) we studied the degree of mixing between star-forming and Seyfert narrow-line region (NLR) emission in the SDSS Seyfert spectra.  In this study we present further results, including the parameter estimates for the metallicity, pressure and ionization parameter.  We additionally expand our analysis to the SDSS star-forming galaxies to contextualize the results of the mixing analysis.

\section{Method} \label{sec:Method}

Our approach uses the analysis tool NebulaBayes \citep{Thomas_2018_NebulaBayes}, which performs Bayesian parameter estimation by comparing observed emission-line fluxes and errors with grids of theoretical fluxes.

\subsection{Observational data} \label{sec:Obs}
We use a similar sample to the sample used in Paper~1.  The parent sample is SDSS DR7 \citep{Abazajian_2009_SDSS7}, and we use the emission line fluxes and stellar masses measured by the MPA-JHU group\footnote{\url{www.mpa-garching.mpg.de/SDSS/DR7/}\\ \url{http://home.strw.leidenuniv.nl/~jarle/SDSS/}} \citep{Tremonti_2004_MZ}.  We apply the empirical scalings to the line flux errors that were derived by that group using duplicate observations.  The following cuts were applied to the sample:
\begin{enumerate}
	\item  We require a signal-to-noise ratio of 5 for the \Hb\ line flux.  This cut is only applied to \Hb\ because cuts on other lines may introduce biases -- in particular, a cut on \OIII\ may bias against high-metallicity objects with weak \OIII.
	\item  We include objects having slightly negative fluxes for other lines, to allow for errors in the continuum subtraction and line fitting -- for a flux $F_i$ with error $E_i$, we require that $F_i > -2.5 E_i$.  This cut is applied separately in the star-forming galaxy and Seyfert galaxy analyses on all lines of the chosen set in each case (Section~\ref{sec:method_apply_NebulaBayes}).
	\item  We take only the first of any duplicate observations
	\item  Galaxies with an unphysical Balmer decrement are excluded, because this is indicative of errors in calibration and continuum subtraction.  We require $F_{\rm H\alpha}/F_{\rm H\beta} > 2.7$.
	\item  We apply standard cuts on the \NII\ and \SII\ optical diagnostic diagrams \citep{Kewley2006_AGN_hosts}.  We include only `star-forming galaxy' and `Seyfert' classifications and deliberately exclude LINERs, composite galaxies and `ambiguous' galaxies that have conflicting classifications on the two diagrams.  Negative fluxes are set to zero for the purposes of the classification.
\end{enumerate}

We did not apply redshift cuts, but systematically consider the effect of different redshift selections in Section~\ref{sec:Discussion}.  After the above cuts there remained 231429 spectra classified as star-forming and 7669 spectra classified as Seyfert~2.  We explore the effects of sample selection in Section~\ref{sec:Discussion}.

\subsection{Model data} \label{sec:Grids}

We use MAPPINGS~V \citep{Sutherland_Dopita_2017_shocks_MAPPINGSV} photoionization models for both the \HII\ regions and NLRs.  All models are plane-parallel, one-dimensional, and dusty, with elemental depletions onto dust grains based on $\log({\rm Fe}_{\rm free} / {\rm Fe}_{\rm total}) = -1.5$ \citep{Jenkins_2009_depletions, Jenkins_2014_depletions}.  Total elemental abundances are set using the oxygen-based standard abundance scaling of \citet{Nicholls_2017_abundance_scaling} and the `local galactic concordance' reference abundances, in which the Solar oxygen abundance is $12 + \log\,{\rm O/H} = 8.76$.  The total pressure in a model is equal to the sum of the gas and radiation pressure such that the total pressure increases as radiation is absorbed in each model step.

For our analysis of the Seyfert galaxies we use a `mixing' grid that combines the photoionization models of \HII\ regions and NLRs.  The mixing grid has four parameters in total: oxygen abundance $12 + \log {\rm O/H}$, NLR ionization parameter $\log U_{\rm NLR}$, gas pressure $\log P/k$, and the fraction of the Balmer flux arising in the NLR as opposed to \HII\ regions, \fN, with the ionizing Seyfert spectrum identical in all models.

Assumptions are employed to reduce the number of free model parameters in the mixing analysis; limiting the number of parameters is necessary because of computational practicalities and the limited number of independent measured line fluxes in each spectrum.  The \HII\ and Seyfert nebulae are assumed to have the same metallicity and gas pressure.  The ionization parameter in the \HII-region component is fixed to $\log U_{\rm HII} = -3.25$, a representative value for high-metallicity galaxies.  In Section~\ref{sec:Discussion} we investigate the sensitivity of the results to the treatment of gas pressure and to the choice of fixed $\log U_{\rm HII}$.

The Seyfert model components use an ionizing spectrum from \citet{Thomas_2016_oxaf} that parameterizes the energy of the ionizing accretion disk emission by its peak energy, \Ep.  In Paper~1 we concluded that values of \Ep\ in the range $40-50$~eV result in plausible distributions of \fN\ measurements.  In the present work we assume \Ep$\,= 45$\,eV.

For our analysis of star-forming galaxies, the \HII-region MAPPINGS grid runs over the three parameters of oxygen abundance, ionization parameter, and gas pressure.  The grid is identical to that described by \citet{Thomas_2018_NebulaBayes}.  A diffuse ionized gas (DIG) component is not included in the models.

\subsection{Application of NebulaBayes} \label{sec:method_apply_NebulaBayes}

In both the \HII\ and mixing analyses we use the Balmer lines \Ha\ and \Hb\ for reddening correction and normalization.  We use the ability of NebulaBayes to deredden spectra to match the computed Balmer decrement at every point in the model grid.  The dereddening method is described in the appendix of \citet{Vogt_2013_HCG_I}.

For the calculation of the likelihood in the mixing analysis we use the emission lines \OII$\,\lambda\lambda\,3726,$3729, [\ion{Ne}{3}]$\,\lambda\,3869$, \OIII$\,\lambda\,4363$, \Hb, \OIII$\,\lambda\,5007$, \ion{He}{1}$\,\lambda\,5876$, [\ion{O}{1}]$\,\lambda\,6300$, \Ha, \NII$\,\lambda\,6583$, and \SII$\,\lambda\lambda\,6716,$ 6731.  These lines include strong optical lines as well as higher-excitation lines that are characteristic of Seyfert nuclei.  The fluxes are normalized to \Hb\ for comparison with the models in the calculation of the likelihood, but we know that other line ratios may effectively constrain parameters of interest.  Hence, we use the NebulaBayes `line ratio prior' feature \citep[described in the appendix of][]{Thomas_2018_NebulaBayes} to apply priors using two diagnostic ratios.  These are \NIIOII\ \citep[a sensitive metallicity diagnostic for metallicites above half Solar;][]{Kewley_and_Dopita_2002_Z}, which slightly reduces the scatter in the inferred O/H, and \SII\,$\lambda 6731$ / \SII\,$\lambda 6716$, an electron density diagnostic \cite[e.g.][]{Dopita_Sutherland_2003_ISM_book, Osterbrock_Ferland_2006_Gas_astro} that improves our constraints on the gas pressure.  The priors are calculated by comparing observed and model ratios over the entire parameter space, similarly to the calculation of the likelihood.

To fit the star-forming galaxies, we relied only on diagnostic line ratios by using the NebulaBayes `line ratio prior' as the likelihood (instead of the usual method of likelihood calculation).  In particular, we used the \NIIOII\ ratio (N2O2) to constrain O/H, the \OIIIOII\ ratio to constrain the ionization parameter, and the \SII\,$\lambda 6731$ / \SII\,$\lambda 6716$ ratio to constrain the pressure, with the N2O2 constraint weighted three times more heavily than each of the other two constraints.  This approach recognizes the sensitivity of the N2O2 ratio to metallicity, and was necessary to achieve a smooth metallicity distribution over the entire range in O/H of more than a decade.  This analysis used a uniform prior (in the logarithmic space of each parameter).

It is not meaningful to compare a model emission line flux to a negative measured flux, so we set any negative line flux $f_i$ with error $e_i$ to $f_i = +0.05 e_i$ in the NebulaBayes analysis.  This produces effectively the same constraint as $f_i = 0$ but satisfies the requirement of the code for positive input fluxes.

NebulaBayes allows the inclusion of a relative error on the model grid fluxes, for which we used a value of 0.35 (as a proportion of each model flux; see Section~2.2 of \citealt{Thomas_2018_NebulaBayes}).  This `grid error' is used in both likelihood and prior calculations.  An error on predicted fluxes of 35\% may often be larger than errors on observed fluxes, but this value is necessary to `smooth out' the constraints and produce a smooth metallicity distribution for the star-forming galaxies.  We explore why the large `grid error' is necessary in Section~\ref{sec:HII_eval}.  For the star-forming galaxy analysis we used cubic interpolation of the model grids.

\section{Results} \label{sec:Results}

\subsection{Parameter estimates} \label{sec:param_estimates}
\begin{figure*}[hbtp]
	\centering
	\includegraphics[width=0.97\textwidth]{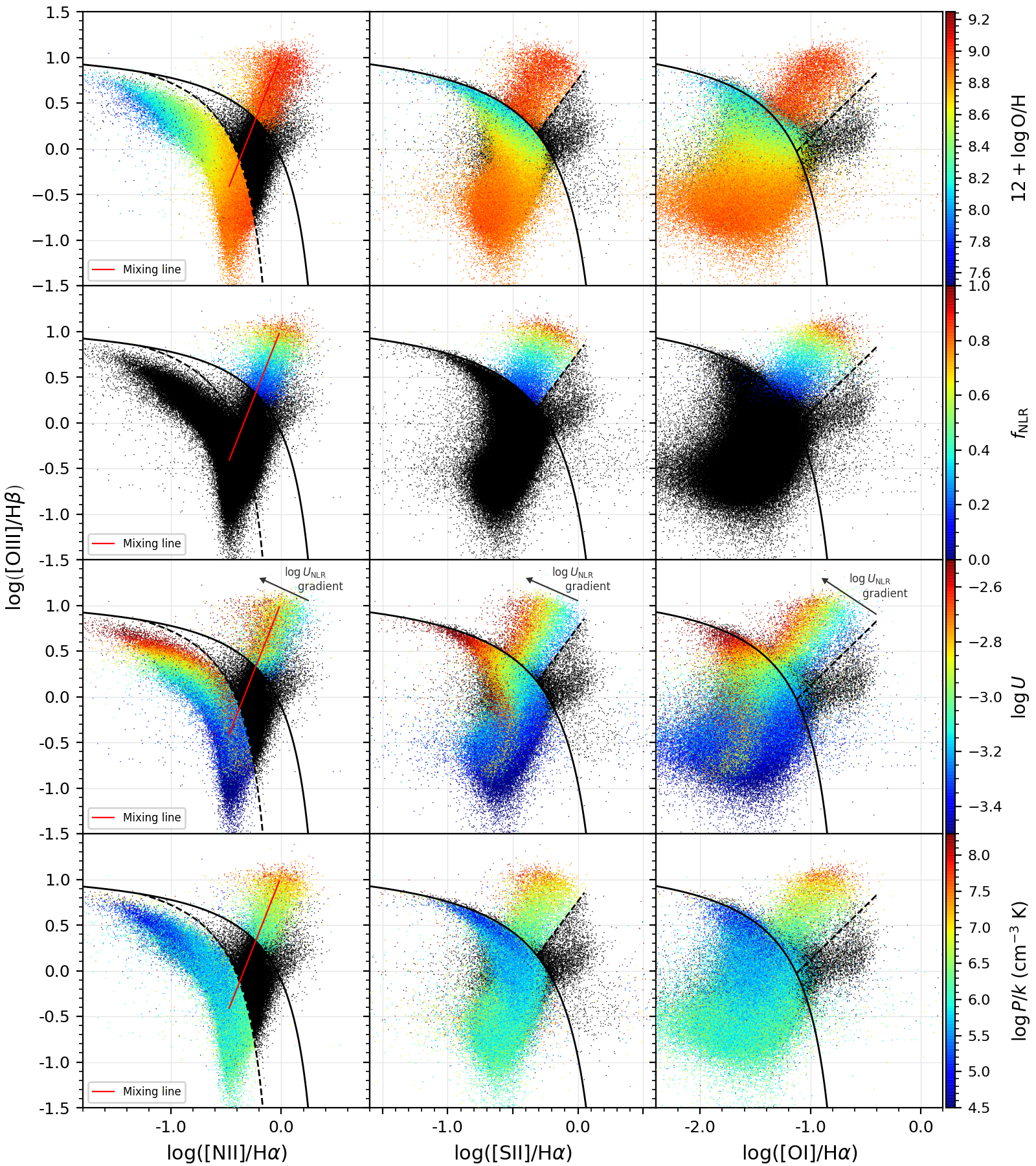}
	\caption{Optical diagnostic diagrams for SDSS emission-line galaxies with data colored by the NebulaBayes parameter estimates.  Only spectra with `Seyfert' or `\HII' classifications were analyzed; other data is shown in black below any colored data.  The red line in each panel of the leftmost column is an approximate `mixing line' designed to trace the upper part of the locus of data connecting the \HII\ and AGN regions (see Paper~1).  The Seyfert ionization parameter estimates are for the NLR component of the mixed models ($\log U_{\rm NLR}$); the \HII\ components had a fixed $\log U_{\rm HII} = -3.25$.  From the top row to the bottom, the diagrams show that Seyfert NLRs are generally high-metallicity, highly contaminated by \HII\ emission, systematically varying in $\log U_{\rm NLR}$, and are characterized by a high gas pressure.
		\label{fig:BPTVO}}
\end{figure*}

Figure~\ref{fig:BPTVO} presents optical diagnostic diagrams \citep{BPT1981, Veilleux_1987_LineRatios} showing the SDSS data colored by the parameter estimates.
 
There is systematic variation in measurements of the ionization parameter $U$ in both the AGN and star-forming sequences.  The gradient for Seyferts is perpendicular to the mixing line, with lower $U$ inferred for data closer to the LINER region \citep[a region associated with low $U$ in the case of photoionized sources, e.g.][]{Binette_1994_pAGB_ionization}.  The anticorrelation of $U$ with metallicity in star-forming galaxies has been previously noted \citep[e.g.][]{Nagao_2006_HII_Z, Dopita_2006_SB_SED_III}.

\begin{figure}[hbtp]
	\centering
	\includegraphics[width=0.46\textwidth]{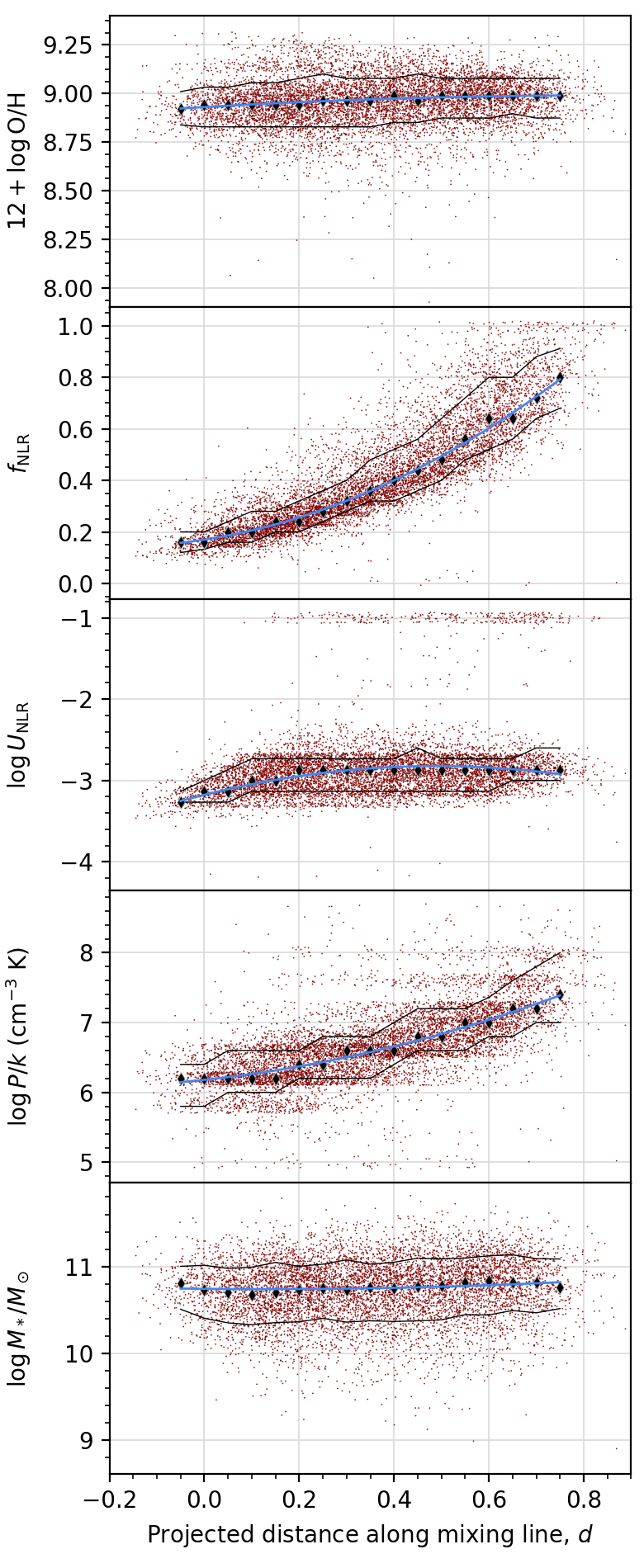}
	\caption{Derived values for SDSS Seyfert~2 galaxies plotted against $d$, the dimensionless projected distance along the mixing line on the log-log BPT diagram (red lines in Figure~\ref{fig:BPTVO}).  The value $d = 0$ corresponds to the intersection of the mixing line and the \citet{Kewley2001} extreme starburst line.
	    \label{fig:mixing}}
\end{figure}

Figure~\ref{fig:mixing} presents the parameter estimates for the Seyfert spectra as a function of projected distance along the mixing sequence, $d$.  Here the mixing sequence (defined in Paper~1) is a straight line on the log-log optical diagnostic diagram designed to follow the upper section of the locus of data that connects the \HII\ and AGN regions.  The panel for \fN\ is the same as the panel for \Ep$\, = 45$\,eV in Figure~2 of Paper~1.

From Figure~\ref{fig:mixing} we note the following:
\begin{enumerate}
	\item Metallicity:  The high $12 + \log {\rm O/H}$ measurements (medians of $1.4 - 1.7$~solar) are consistent with the `mixing' with the high-metallicity end of the \HII-region sequence (Figure~\ref{fig:BPTVO}).  The ${\sim}0.1$\,dex increase in median metallicities with $d$ is presumably related to a systematic offset between inferred \HII-region and NLR metallicities (Section~\ref{sec:systematics}).
	\item Mixing fraction:  The \fN\ estimates show that the Seyfert-classified SDSS spectra are highly contaminated by \HII-region emission, and \fN\ increases smoothly along the mixing sequence (the \fN\ results are the subject of Paper~1).
	\item Ionization parameter:  The measured $U_{\rm NLR}$ is typically in the tight range $-3.3 < \log U_{\rm NLR} < -2.7$, and shows little variation as a function of $d$.  The gradient in $\log U_{\rm NLR}$ in Figure~\ref{fig:BPTVO} is not evident in Figure~\ref{fig:mixing} because the gradient is approximately orthogonal to the mixing line.  The lower inferred $U$ at low \fN\ (low $d$) may be due to LINER or DIG contamination.
	\item Pressure:  The median inferred $\log P/k$ values increase by approximately 1.2~dex along the mixing sequence.
\end{enumerate}

The gas pressure is measured to be higher in galaxies with emission dominated by Seyfert excitation than in galaxies of similar masses that are dominated by star-forming excitation.  This is evident in the differences between the top and bottom ends of the mixing sequence in Figure~\ref{fig:BPTVO} and Figure~\ref{fig:mixing}.  The higher pressures in NLRs are presumably caused by the effects of radiation pressure \citep[e.g.][]{Dopita_2002_NLRs, Groves_2004_NLR_models_II} and potentially also the ram pressure of AGN-induced outflows.  The single pressure parameter in our mixing analysis is effectively a `mean pressure', which naturally increases as \fN\ increases and higher-pressure NLRs dominate the emission.

\subsection{Mass-metallicity relations} \label{sec:results_SDSS}

\begin{figure*}[hbtp]
	\centering
	\includegraphics[width=0.95\textwidth]{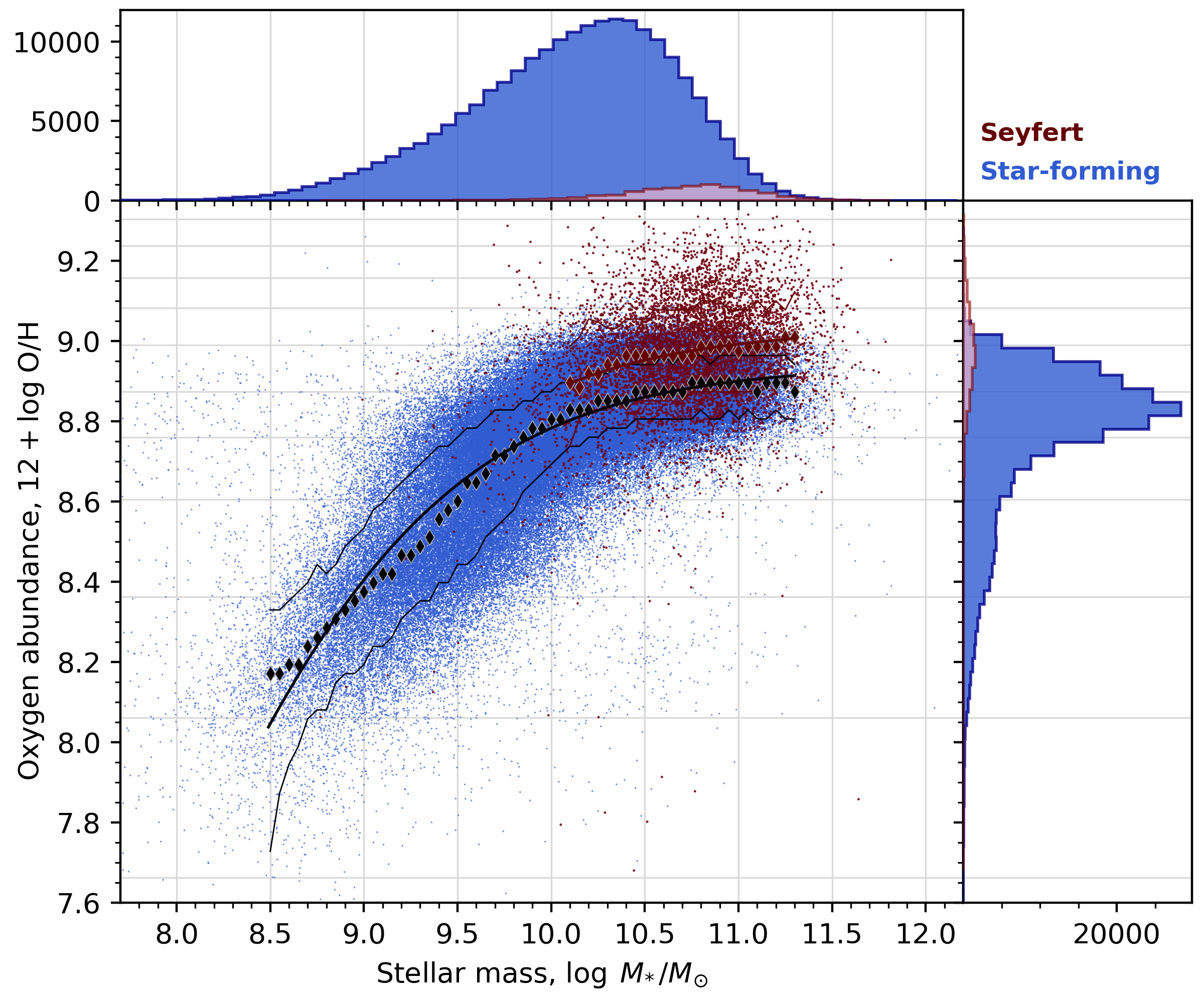}
	\caption{The SDSS galaxy mass-metallicity relations resulting from our inferred oxygen abundances.  Horizontal gridlines show the photoionization model metallicities used for the \HII, NLR and mixed grids.  Marginal histograms show the distributions of stellar mass and oxygen abundance for the star-forming galaxies (blue) and the AGN (red).  The results reproduce the well-known relationship between galaxy stellar mass and gas-phase oxygen abundance in SDSS star-forming galaxies \citep[e.g.][]{Tremonti_2004_MZ}.  The Seyfert~2 galaxies follow a locus similar to the upper end of the star-forming galaxy sequence.  The offset between the star-forming and active galaxy relations is explored in Section~\ref{sec:systematics}.
		\label{fig:MZ_results}}
\end{figure*}

Figure~\ref{fig:MZ_results} presents the mass-metallicity relation resulting from our oxygen abundance measurements for both the \HII- and Seyfert-classified spectra.  Our derived AGN mass-metallicity relation is similar to that of star-forming galaxies, albeit covering only the high-mass end of the relation.

We fit the following functional form \citep{Moustakas_2011_MZ_evolution, Zahid_2014_MZ_II} to the mass-metallicity relations:  
\begin{equation}
\theta(M_\star) = \theta_0 - \log \left[ 1 + \left( \frac{M_\star}{M_0} \right )^{-\gamma\,} \right]  \label{eq:MZ}
\end{equation}
where $\theta = 12 + \log {\rm O/H}$ is the oxygen abundance and $M_\star$ is the galaxy stellar mass.  The three parameters are $\gamma$, the power-law slope of the relation at low stellar masses; $M_0$, a characteristic stellar mass at which the mass-metallicity relation begins to flatten; and $\theta_0$, the asymptotic oxygen abundance at high stellar masses.  Equation~\ref{eq:MZ} increases monotonically with $M_\star$.

Our approach is to firstly fit the mass-metallicity sequence of the star-forming galaxies, and then apply the resulting values of $\gamma$ and $M_0$ in the fit to the AGN sequence, such that only $\theta_0$ may vary.  This approach is necessary because the AGN sequence does not extend to sufficiently low stellar masses to effectively constrain $\gamma$ and $M_0$.  The fits were performed over the ranges $8.5 < \log M_* / M_\odot < 11.3$ and $10.1 < \log M_* / M_\odot < 11.3$ for the star-forming and active galaxies, respectively.

Our derived AGN mass-metallicity relation is similar in shape to the relation for star-forming galaxies, albeit covering only the high-mass end of the relation.  We detect an increase in AGN metallicity with galaxy mass, with the median oxygen abundance increasing by ${\sim}0.1$~dex over the fitted range of 1.3\,dex in stellar mass.  This is consistent with the observation of \citet{Groves_2006_SDSS_AGN} that \NII\ emission generally increases with the stellar mass of SDSS active galaxies, interpreted by the authors as an increase in metallicity with mass.

The scatter about the derived mass-metallicity relations, characterized by the median absolute residual, is 0.07~dex in O/H for both mass-metallicity relations.  The NebulaBayes errors on individual metallicity measurements are generally more than three times larger than the residuals; these errors effectively include both statistical and systematic unceratinties because of the 35\% error used on the grid fluxes (Section~\ref{sec:method_apply_NebulaBayes}; Section~\ref{sec:HII_eval}).

\vspace{0.5cm}
\subsection{Comparison to the literature}
\label{sec:MZ_compare}

\begin{figure}[hbtp]
	\centering
	\includegraphics[width=0.48\textwidth]{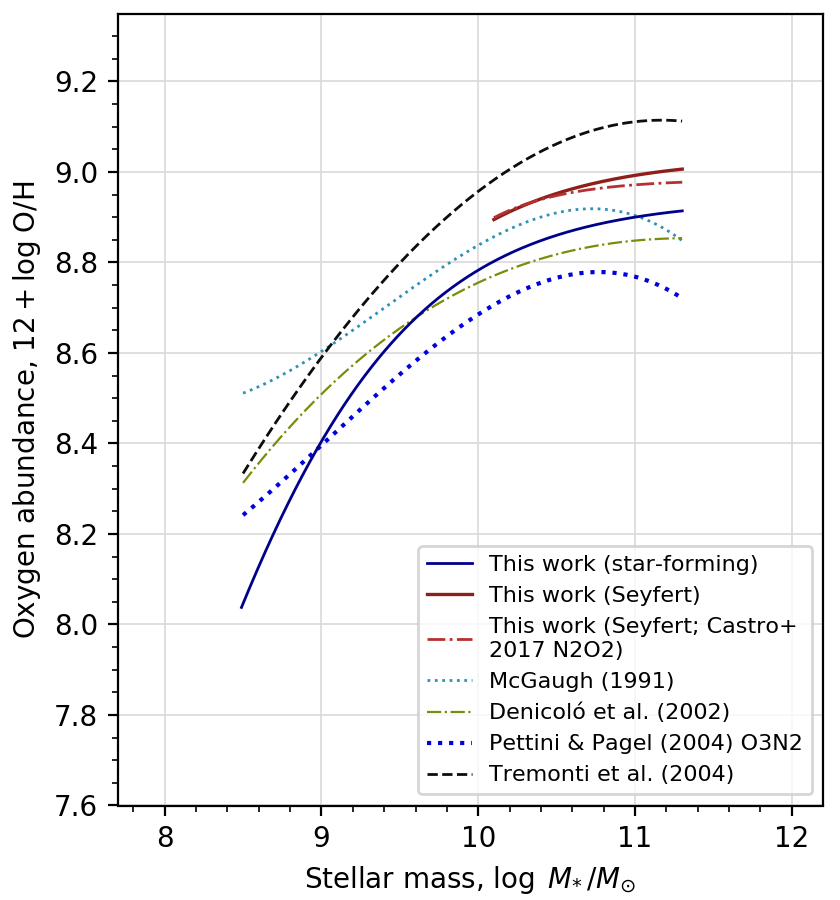}
	\caption{Comparison of our inferred mass-metallicity relations with those derived using other metallicity estimates.  The \citet{Castro_2017_AGN_Z} N2O2 diagnostic was applied to our Seyfert sample to produce the corresponding relation (after correcting for a different assumed solar abundance).  The relations for the calibrations of \citet{McGaugh_1991_Z}, \citet{Denicolo_2002_N2}, \citet{Pettini_and_Pagel_2004_Z_diagnostic} and \citet{Tremonti_2004_MZ} are as reported in Table~2 of \citet{Kewley_Ellison_2008_Z}.
	 \label{fig:MZ_comparison}}  
\end{figure}

Figure~\ref{fig:MZ_comparison} compares our results to some other calculated mass-metallicity relations.  We apply the recent N2O2 metallicity calibration of \citet{Castro_2017_AGN_Z} to our Seyfert sample and include the resulting relation in the figure.  We also show a selection of relations calculated by \citet{Kewley_Ellison_2008_Z} using various diagnostics and a similar star-forming galaxy sample to our own.  The well-known systematic offsets between different metallicity calibrations \citep[e.g.][]{Kewley_Ellison_2008_Z} are apparent.  We compare our results to those of \citet{Tremonti_2004_MZ} in more detail in Section~\ref{sec:method_comparison}.

The mass-metallicity relation we derive using the \citet{Castro_2017_AGN_Z} diagnostic is very similar to that produced by our NebulaBayes analysis in its shape, scatter, and absolute oxygen abundance values (after correcting for the 0.07~dex difference in assumed solar oxygen abundance between the \citet{Castro_2017_AGN_Z} models and our own).  The \citet{Castro_2017_AGN_Z} photoionization models neglect mixing between \HII-region and NLR emission, as well as variations in pressure.  Despite these simplifications, the \citet{Castro_2017_AGN_Z} diagnostic has produced an almost identical Seyfert mass-metallicity relation to our more sophisticated analysis, which suggests that N2O2 is a very robust metallicity diagnostic, both for Seyfert galaxies and in general.

\section{Further analysis and discussion} \label{sec:Discussion}
We confirm the results of \citet{Groves_2006_SDSS_AGN} and others that sub-solar AGN are rare in the SDSS sample.  Low-metallicity AGN should lie between the `wings' of the BPT `butterfly diagram' \citep[e.g.][]{Kawasaki_2017_SDSS_lowZ_AGN}; in our sample objects in this region appear to have a combination of high ionization parameter and relatively low metallicity (Figure~\ref{fig:BPTVO}).

\subsection{What causes the systematic offset between star-forming and active galaxies?}
\label{sec:systematics}
There is a systematic offset in oxygen abundance of $0.09$~dex between the star-forming and active galaxies in Figure~\ref{fig:MZ_results}.  We do not expect that the gas-phase metallicities of star-forming galaxies and AGN should be different for samples of galaxies at the same mass, although there is at least one mechanism that speculatively may produce abundance enhancements in active nuclei -- star formation with a top-heavy IMF in the accretion disk \citep[e.g.][]{ Nayakshin_2005_SgrAstar_disk_SF, Collin_2008_Accretion_disc_SF, Bartko_2010_SgrAstar_topheavy_IMF}.  In this section we explore potential methodological causes of the offset by reanalyzing the data in a series of experiments.  The \HII\ region and NLR models used the same photoionization code (MAPPINGS~V) and abundances, so these could not have contributed to the offset.  The reanalyses involved varying the lines and mixing grids used by NebulaBayes, as well as varying the selection of the sample.  The experiments are summarized in Table~\ref{table:offsets}.

\begin{deluxetable*}{lccccccccccccccc}
	\centering
	\tabletypesize{\scriptsize}
	\tablewidth{12cm}
	\tablecaption{The measured mass-metallicity relation parameters $\gamma$, M$_0$ and $\theta_0$ for different choices in the analysis, along with $\theta_0$ offsets between corresponding star-forming galaxy and AGN mass-metallicity relations. \label{table:offsets}}
	\tablehead{
		\multicolumn{1}{l}{\#} &
		\multicolumn{1}{l}{Type$^{a}$} &
		\multicolumn{1}{c}{Lines} &
		\multicolumn{1}{c}{$\log U_{\rm HII}$} &
		\multicolumn{1}{c}{$\log P_{\rm HII}\,/\,k\,^{c}$} &
		\multicolumn{2}{c}{redshift, $z$} &
		\multicolumn{2}{c}{EW(H$\beta$)} &
		\multicolumn{2}{c}{Low-M$_\star$ slope, $\gamma$} &
		\multicolumn{2}{c}{Turnover, M$_0$} &
		\multicolumn{3}{c}{Asymptotic O/H, $\theta_0$} \\
		\multicolumn{1}{l}{} &  
		\multicolumn{1}{l}{} &  
		\multicolumn{1}{c}{\& line} &
		\multicolumn{1}{c}{} &  
		\multicolumn{1}{c}{(cm$^{-3}$ K)} &  %
		\multicolumn{1}{c}{} &  
		\multicolumn{1}{c}{} &  
		\multicolumn{2}{c}{(\AA)} &  
		\multicolumn{1}{c}{} &
		\multicolumn{1}{c}{} &
		\multicolumn{2}{c}{($\log$ M$_\star / $M$_\odot$)} &
		\multicolumn{3}{c}{(12 + $\log$ O/H)} \\  
		\multicolumn{1}{l}{} &  
		\multicolumn{1}{l}{} &  
		\multicolumn{1}{c}{ratios$^{b}$} &
		\multicolumn{1}{c}{} &  
		\multicolumn{1}{c}{} &  %
		\multicolumn{1}{c}{min} &
		\multicolumn{1}{c}{max} &
		\multicolumn{1}{c}{min} &
		\multicolumn{1}{c}{max} &
		\multicolumn{1}{c}{value} &
		\multicolumn{1}{c}{error} &
		\multicolumn{1}{c}{value} &
		\multicolumn{1}{c}{error} &
		\multicolumn{1}{c}{value} &
		\multicolumn{1}{c}{error} &
		\multicolumn{1}{c}{offset$^{d}$}
	}
	\startdata
	\decimals
	1$^{e}$ &  SF  &  A       &   free &  free          &   -    &   -    &   -   &   -   &  0.62 &  0.03 &  8.76 &  0.02 &  8.94 &  0.009 &  -    \\
	2       &  AGN &  B       &  -3.50 &  $P_{\rm NLR}$ &   -    &   -    &   -   &   -   &  "    &  "    &  "    &  "    &  9.01 &  0.002 &  0.07 \\
	3$^{e}$ &  AGN &  B       &  -3.25 &  $P_{\rm NLR}$ &   -    &   -    &   -   &   -   &  "    &  "    &  "    &  "    &  9.03 &  0.002 &  0.09 \\
	4       &  AGN &  B       &  -3.00 &  $P_{\rm NLR}$ &   -    &   -    &   -   &   -   &  "    &  "    &  "    &  "    &  9.05 &  0.003 &  0.11 \\
	5       &  AGN &  B       &  -3.25 &  5.80          &   -    &   -    &   -   &   -   &  "    &  "    &  "    &  "    &  9.02 &  0.002 &  0.08 \\
	6       &  AGN &  B       &  -3.25 &  6.20          &   -    &   -    &   -   &   -   &  "    &  "    &  "    &  "    &  9.03 &  0.002 &  0.09 \\
	7       &  AGN &  B       &  -3.25 &  6.60          &   -    &   -    &   -   &   -   &  "    &  "    &  "    &  "    &  9.04 &  0.002 &  0.10 \\
	8$^{f}$ &  AGN &  A$^{f}$ &  -3.25 &  $P_{\rm NLR}$ &   -    &   -    &   -   &   -   &  "    &  "    &  "    &  "    &  9.00 &  0.003 &  0.06 \\
	\hline                                                                                                                                
	9       &  SF  &  A       &   free &  free          &  0.000 &  0.050 &   -   &   -   &  0.63 &  0.02 &  8.82 &  0.02 &  8.99 &  0.009 &  -    \\
	10       &  AGN &  B       &  -3.25 &  $P_{\rm NLR}$ &  0.000 &  0.050 &   -   &   -   &  "    &  "    &  "    &  "    &  9.06 &  0.006 &  0.07 \\
	11       &  SF  &  A       &   free &  free          &  0.050 &  0.075 &   -   &   -   &  0.54 &  0.04 &  8.85 &  0.04 &  8.99 &  0.02  &  -    \\
	12       &  AGN &  B       &  -3.25 &  $P_{\rm NLR}$ &  0.050 &  0.075 &   -   &   -   &  "    &  "    &  "    &  "    &  9.08 &  0.005 &  0.08 \\
	13       &  SF  &  A       &   free &  free          &  0.075 &  0.100 &   -   &   -   &  0.68 &  0.05 &  8.98 &  0.03 &  8.95 &  0.01  &  -    \\
	14       &  AGN &  B       &  -3.25 &  $P_{\rm NLR}$ &  0.075 &  0.100 &   -   &   -   &  "    &  "    &  "    &  "    &  9.05 &  0.006 &  0.10 \\
	15       &  SF  &  A       &   free &  free          &  0.100 &  0.200 &   -   &   -   &  0.76 &  0.05 &  9.09 &  0.02 &  8.93 &  0.01  &  -    \\
	16       &  AGN &  B       &  -3.25 &  $P_{\rm NLR}$ &  0.100 &  0.200 &   -   &   -   &  "    &  "    &  "    &  "    &  9.03 &  0.005 &  0.10 \\
	\hline                                                                                                                                   
	17       &  SF  &  A       &   free &  free          &   -    &   -    &  -30  &  -10  &  0.48 &  0.03 &  8.81 &  0.04 &  8.98 &  0.02  &  -    \\
	18       &  AGN &  B       &  -3.25 &  $P_{\rm NLR}$ &   -    &   -    &  -30  &  -10  &  "    &  "    &  "    &  "    &  9.08 &  0.01  &  0.10 \\
	19       &  SF  &  A       &   free &  free          &   -    &   -    &  -10  &   -6  &  0.66 &  0.03 &  8.78 &  0.02 &  8.93 &  0.007 &  -    \\
	20       &  AGN &  B       &  -3.25 &  $P_{\rm NLR}$ &   -    &   -    &  -10  &   -6  &  "    &  "    &  "    &  "    &  9.04 &  0.005 &  0.11 \\
	21       &  SF  &  A       &   free &  free          &   -    &   -    &   -6  &   -3  &  0.68 &  0.02 &  8.76 &  0.01 &  8.93 &  0.006 &  -    \\
	22       &  AGN &  B       &  -3.25 &  $P_{\rm NLR}$ &   -    &   -    &   -6  &   -3  &  "    &  "    &  "    &  "    &  9.03 &  0.004 &  0.10 \\
	23       &  SF  &  A       &   free &  free          &   -    &   -    &   -3  &    0  &  0.88 &  0.03 &  8.71 &  0.02 &  8.88 &  0.004 &  -    \\
	24       &  AGN &  B       &  -3.25 &  $P_{\rm NLR}$ &   -    &   -    &   -3  &    0  &  "    &  "    &  "    &  "    &  8.97 &  0.004 &  0.09 \\[0.1cm]
	\enddata
	\tablenotetext{a}{``SF'':  An analysis of the star-forming galaxies using an HII-region model grid; ``AGN'':  An anlaysis of AGN using an HII-NLR mixing grid}
	\tablenotetext{b}{The line fluxes and flux ratios that were compared between the observations and models to constrain the parameters for each galaxy:\\
		A: N2O2 (triply weighted), O3O2, \SII$\,\lambda\,6731 / 6716$\\
		B: \OII$\,\lambda\lambda\,3726,$3729, [\ion{Ne}{3}]$\,\lambda\,3869$, \OIII$\,\lambda\,4363$, \OIII$\,\lambda\,5007$, \ion{He}{1}$\,\lambda\,5876$, [\ion{O}{1}]$\,\lambda\,6300$, 	\NII$\,\lambda\,6583$, and \SII$\,\lambda\lambda\,6716,6731$, with priors on N2O2 and \SII$\,\lambda\,6731 / 6716$.}
	\tablenotetext{c}{``free'':  The HII-region pressure varied as its own parameter in the SF galaxy analyses.  A value:  The pressure of the HII-region component of the mixed models was set to this value, and the pressure of the NLR component varied as its own parameter.  ``$P_{\rm NLR}$'':  The pressure of the HII-region component of the mixed models was constrained to be equal to the pressure of the NLR component.}
	\tablenotetext{d}{The difference between the fitted $\theta_0$ values of the AGN analysis and the corresponding SF analysis, $\theta_{0, {\rm AGN}} - \theta_{0, {\rm SF}}$.}
	\tablenotetext{e}{Rows 1 and 3 are the ``preferred'' star-forming galaxy and AGN analyses, for which the results are presented in Figure~\ref{fig:BPTVO}, Figure~\ref{fig:mixing}, Figure~\ref{fig:MZ_results}, and Figure~\ref{fig:MZ_comparison}.}
	\tablenotetext{f}{Row 8 presents a HII-NLR mixing analysis for AGN using the same lines/ratios as the SF analyses (`A').}
\end{deluxetable*}

\subsubsection{HII region ionization parameter}
\label{sec:HII_log_U}
The ionization parameter of the \HII\ region component of the mixed models was set to $\log U_{\rm HII} = -3.25$, a value chosen to be representative of high-metallicity galaxies.  The choice of fixed $U_{\rm HII}$ should affect the inferred AGN metallicities because measurements of ionization parameter and metallicity are interdependent, because $U$ and O/H are observed to be anticorrelated (Figure~\ref{fig:BPTVO}), and because the \HII-region emission contribution is significant for most SDSS AGN.
	
Rows 2 and 4 of Table~\ref{table:offsets} present the results of varying $U_{\rm HII}$ by 0.25~dex below and above the assumed value in row 3.  The inferred AGN metallicities (and hence $\theta_0$ estimates) are found to increase with $U_{\rm HII}$.  It appears that the $\theta_0$ offset could possibly be eliminated by setting an unphysically low value of $U_{\rm HII}$.  Nevertheless, we have determined that the choice of the fixed value of $U_{\rm HII}$ within the plausible range of values is not a significant contributor to the $\theta_0$ offset.

\subsubsection{Treatment of pressure}
\label{sec:systematics:pressure}
In the mixing analysis only one parameter was used to account for varying gas pressure between galaxies.  This approach assumes that the \HII\ region and NLR pressures are equal and the parameter therefore measures an `effective' pressure.

To test the sensitivity to this particular choice of pressure treatment, we performed the experiments presented in rows 5, 6 and 7 of Table~\ref{table:offsets}.  In these experiments we fixed the pressure of the HII-region contribution to the mixed models, such that the free pressure parameter was for the gas pressure in the NLR contribution only.  Inspection of the pressure results for these experiments showed that the distributions of inferred NLR pressures were implausible (not smooth and with an implausibly large range).  The mass-metallicity relation offsets remain of a similar size with this different pressure treatment.  We therefore conclude that the treatment of pressure in the models is unlikely to make a significant contribution to the offset.

\subsubsection{Choices of lines, line ratios and weightings}
\label{sec:systematics:lines}
The most important choices for an analysis with NebulaBayes are in the selection of emission lines, emission line ratios and corresponding weightings for the comparison between models and observations.  The star-forming galaxy and mixing analyses were very different in these choices (Section~\ref{sec:method_apply_NebulaBayes}).  In particular, the star-forming galaxy analysis relied purely on diagnostic line ratios and the mixing analysis took advantage of high-excitation lines that are only observed in AGN.
	
Row 8 of Table~\ref{table:offsets} presents the results of an experiment in which our preferred analysis (row 3) was repeated, but using the lines, line ratios and weightings that were applied in the star-forming galaxy analysis.  This approach produced a reduction in the $\theta_0$ offset from 0.09 to 0.06, indicating that the treatment of lines, line ratios and weightings does substantially contribute to the $\theta_0$ offset.

\subsubsection{Redshift cuts and aperture effects}
\label{sec:systematics:z}
It is possible that aperture effects and changes in sample properties with redshift may have systematically affected our analysis.  To explore these effects, we considered subsamples with different ranges of redshift.  Rows 9 to 16 of Table~\ref{table:offsets} present the results of fitting the star-forming galaxy and AGN mass-metallicity relations for galaxies in four different redshift bins.  The subsamples have $\theta_0$ offsets of 0.07-0.10~dex, with the offset increasing by ${\sim}0.03$ as a function of redshift over the four bins.  The variation in the offset suggests that aperture and selection effects could make a small contribution to the systematic offset in our preferred analysis, but cannot explain the majority of the offset.

\subsubsection{Omission of DIG in the models}
\label{sec:systematics:DIGS}
Emission from DIG is ubiquitous and contributes ${\sim}10 - 70\%$ of the \Ha\ luminosity in star-forming galaxies, with later (earlier) types showing contributions at the lower (higher) end of this range \citep[e.g.][]{Lacerda_2018_CALIFA_DIG}.  Emission from DIG is associated with lower pressure, lower ionization parameter, and harder ionising spectra than \HII-region emission, and DIG contamination may cause systematic offsets in metallicity measurements \citep{Zhang_2017_MaNGA_DIG, Pilyugin_2018_DIG_contamination}.  We expect the mixing analysis to be less sensitive to DIG contamination because NLRs in Seyfert-classified objects are likely far more luminous than DIG, and the NLR models include DIG-like emission from the partially ionized zone.

To explore potential systematic effects of DIG contamination on the analysis, we considered subsamples defined to have different ranges of the equivalent width (EW) of the \Hb\ emission line.  The Balmer line EW is used as an indicator of DIG emission, with relatively small (large) EW associated with higher (lower) DIG contributions to the line emission \citep[e.g.][]{Cid_Fernandes_2011_WHAN, Belfiore_2016_MaNGA_LIERs, Lacerda_2018_CALIFA_DIG}.  Rows 17 to 24 of Table~\ref{table:offsets} present the results of reanalysing the data in each \Hb\ EW bin.  The four subsamples have $\theta_0$ offsets of $0.09-0.11$~dex, and the offset does not vary suggestively as a function of the \Hb\ EW of the bin.  We are therefore confident that the treatment of DIG in the models and DIG contamination of the SDSS spectra do not significantly contribute to the offset.

\subsubsection{Treatment of dust}
\label{sec:systematics:dust}
Dust destruction is not included in our models but does occur in AGN NLRs, as evidenced by observations of the strength of the coronal lines in Seyfert spectra \cite[e.g.][]{Dopita_2015_probing_ENLR_II_S7}.  Dust destruction in NLRs would lead to more free species contributing to line cooling than in models at the same metallicity, resulting in overestimation of metallicities.  It is possible that dust destruction in Seyferts may lead to systematically different reddening compared to star-forming galaxies, but this does not appear to be the case: Seyferts have uncorrelated \fN\ and $A_v$ (Pearson $r = -0.07$); \fN\ and O/H are also uncorrelated ($r = 0.08$).  However, issues of geometry make it difficult to use extinction data to draw conclusions about varying dust destruction between NLRs and \HII\ regions.  Firstly, dust screens may be significantly physically separated from the relevant nebulae.  Secondly, the observed extinction depends on the relative geometry of sources and dust \cite[e.g.][]{Calzetti_1994_extinction}, with NLRs having strongly differing geometries and extents to \HII\ regions.

\subsubsection{The ionizing spectra}
\label{sec:systematics:spectra}
Ionizing stellar spectra are uncertain \citep[e.g.][]{Morisset_2004_ionizing_stellar_spectra} because of model sensitivity to properties such as stellar rotation, helium dredge-up, mass-loss rates, and binarity, and additional issues with stellar track and stellar atmosphere models such as poor parameter-space coverage, interpolation, and inconsistent abundance sets.  The ionizing spectra in NLRs are also uncertain, although our \fN\ parameter may partially compensate for variations in the ionizing Seyfert spectrum.  If the offset primarily arises due to issues with the assumed ionizing spectra, it is unclear whether the problem is due to the stellar spectra, AGN spectra, or a combination of the two.

\subsubsection{Summary}
\label{sec:systematics:summary}

The major cause of the $0.09$~dex systematic offset in measured oxygen abundance between the star-forming and active galaxies has not been identified.  Our analyses suggest that the offset is not primarily due to the treatment of pressure or the \HII-region ionization parameter in the mixing analysis, to a sample selection effect, or to contamination of the observed spectra by DIG.  We determined that a non-negligible contribution to the offset (${\sim}0.3$~dex) was made by the differences between the choices of emission lines in the two analyses.  It remains unclear what causes the remainder of the offset.

The unexplained ${\sim} 0.06$~dex offset may be due to real physical differences between the observed \HII\ regions and NLRs that we do not account for in the models, including issues of geometry and the treatment of dust.  We also suspect that the highly uncertain ionizing spectra make significant contributions to the offset.

\subsection{Comparison of method with \citet{Tremonti_2004_MZ}}
\label{sec:method_comparison}

We compare our results to those of \citet{Tremonti_2004_MZ}, whose mass-metallicity relation is a classic result in the literature and who followed a methodology similar to our own \citep{Brinchmann_2004_SDSS_Z}. The shape of our star-forming galaxy mass-metallicity relation resembles that derived by \citet{Tremonti_2004_MZ} (Figure~\ref{fig:MZ_comparison}; \citet{Tremonti_2004_MZ} fitted a cubic polynomial).  The scatter is also similar, with \citet{Tremonti_2004_MZ} reporting a 1$\sigma$ spread in O/H of approximately $\pm 0.1$~dex ($\pm$0.07~dex at high galaxy masses), which compares to a median absolute residual of $\pm$0.07~dex for both our star-forming and active galaxy relations.  The similarity in the scatter is consistent with the similarities between the methods of \citet{Tremonti_2004_MZ} and our methods.

There is a systematic O/H offset between the star-forming galaxy mass-metallicity relation of \citet{Tremonti_2004_MZ} and the relation derived in this work (Figure~\ref{fig:MZ_comparison}), which we suspect is due predominantly to differences in the modeling.  \citet{Tremonti_2004_MZ} used theoretical models by \citet{Charlot_and_Longhetti_2001_SF_lines} that had been calibrated to observed line ratios in nearby galaxies, whereas our models are purely theoretical.  Another difference is that we fitted diagnostic line ratios in our star-forming galaxy analysis, whereas \citet{Tremonti_2004_MZ} used line fluxes instead of flux ratios (similarly to our AGN analysis).  However, the choice of lines, ratios and priors did not contribute most of the offset between our star-forming and active galaxy relations (Section~\ref{sec:systematics:lines}), so we suspect the larger offset considered here is also mostly independent of these choices.

\subsection{Evaluation of \HII-region models} \label{sec:HII_eval}
We have compared the observed and best-fit model fluxes for the star-forming galaxies in bins of measured O/H.  The comparison gives insight into why the star-forming galaxy analysis required a different approach to the Seyfert analysis (Section~\ref{sec:method_apply_NebulaBayes}).

Considering measurements of a given line for all galaxies in an O/H bin, we find typical offsets of ${\sim}20 - 30\%$ between the median dereddened observed flux and median predicted flux.  The offsets are similar to the $0.1 - 0.15$~dex quoted by \cite{Kewley_Ellison_2008_Z} as a common estimate for model uncertainties.  However, an exception is \OIII$\lambda 5007$, which is systematically underpredicted by a factor of ${\sim}3 - 5$ at supersolar metallicities.  It is likely that the difficulties fitting the \HII-region models were primarily attributable to this \OIII\ offset.  The offsets may have many causes (such as those considered in Section~\ref{sec:systematics}), but the underprediction of \OIII\ in particular suggests that the model ionising stellar spectra may be inaccurate (Section~\ref{sec:systematics:spectra}).

In general, the systematic offsets explain why it was necessary to use the NebulaBayes `grid error' input to obtain a smooth star-forming galaxy metallicity distribution (Section~\ref{sec:method_apply_NebulaBayes}).  A consequence of using this parameter is that the derived uncertainties on individual parameter estimates are effectively a combination of both statistical uncertainties and model-related systematic uncertainties, rather than being solely statistical uncertainties.  We note that the grid error also allows for systematic errors in the observations, e.g.\ due to difficulties with continuum subtraction.

\section{Conclusions} \label{sec:Conclusions}
We have used a Bayesian method to analyze the line emission in SDSS AGN and star-forming galaxies, in particular making systematic measurements of the gas-phase metallicities.  The method accounts for mixing between \HII-region and NLR emission in AGN for the first time.

We conclude that the inferred AGN metallicities increase with the galaxy stellar mass, as is the case for star-forming galaxies.  The SDSS AGN are located in high-metallicity and high-mass galaxies, and are rare or difficult to detect in lower-mass galaxies.

The shape of the AGN mass-metallicity relation is consistent with the shape of the high-mass end of the star-forming galaxy relation.  Both our star-forming and active galaxy relations have a median absolute residual (scatter) of 0.07~dex in oxygen abundance.  The zero points of the mass-metallicity relations differ, however, with the inferred oxygen abundances for Seyfert galaxes being systematically 0.09~dex higher than for star-forming galaxies.  We have investigated some potential causes of this offset, and although the difference remains largely unexplained we suspect that the model ionizing spectra make a significant contribution.  Nevertheless, we conclude that our method produces robust relative AGN metallicities.

We look forward to applying similar analyses across large samples of AGN with spectroscopic data that is spatially-resolved, beginning with data from the S7 survey \citep{Dopita_2015_probing_ENLR_II_S7, Thomas_2017_S7_DR2}.

%
%
%
%
%

\acknowledgments
This research was conducted by the Australian Research Council Centre of Excellence for All Sky Astrophysics in 3 Dimensions (ASTRO 3D), through project number CE170100013.  B.G. acknowledges the support of the Australian Research Council as the recipient of a Future Fellowship (FT140101202).  M.D. and R.S. acknowledge support from ARC discovery project \#DP160103631.  This research is supported by an Australian Government Research Training Program (RTP) Scholarship.


\begin{thebibliography}{}
	\expandafter\ifx\csname natexlab\endcsname\relax\def\natexlab#1{#1}\fi
	\providecommand{\url}[1]{\href{#1}{#1}}
	
	\bibitem[{{Abazajian} {et~al.}(2009){Abazajian}, {Adelman-McCarthy},
		{Ag{\"u}eros}, {Allam}, {Allende Prieto}, {An}, {Anderson}, {Anderson},
		{Annis}, {Bahcall}, \& et~al.}]{Abazajian_2009_SDSS7}
	{Abazajian}, K.~N., {Adelman-McCarthy}, J.~K., {Ag{\"u}eros}, M.~A., {et~al.}
	2009, \apjs, 182, 543
	
	\bibitem[{{Baldwin} {et~al.}(1981){Baldwin}, {Phillips}, \&
		{Terlevich}}]{BPT1981}
	{Baldwin}, J.~A., {Phillips}, M.~M., \& {Terlevich}, R. 1981, \pasp, 93, 5
	
	\bibitem[{{Bartko} {et~al.}(2010){Bartko}, {Martins}, {Trippe}, {Fritz},
		{Genzel}, {Ott}, {Eisenhauer}, {Gillessen}, {Paumard}, {Alexander},
		{Dodds-Eden}, {Gerhard}, {Levin}, {Mascetti}, {Nayakshin}, {Perets},
		{Perrin}, {Pfuhl}, {Reid}, {Rouan}, {Zilka}, \&
		{Sternberg}}]{Bartko_2010_SgrAstar_topheavy_IMF}
	{Bartko}, H., {Martins}, F., {Trippe}, S., {et~al.} 2010, \apj, 708, 834
	
	\bibitem[{{Belfiore} {et~al.}(2016){Belfiore}, {Maiolino}, {Maraston},
		{Emsellem}, {Bershady}, {Masters}, {Yan}, {Bizyaev}, {Boquien}, {Brownstein},
		{Bundy}, {Drory}, {Heckman}, {Law}, {Roman-Lopes}, {Pan}, {Stanghellini},
		{Thomas}, {Weijmans}, \& {Westfall}}]{Belfiore_2016_MaNGA_LIERs}
	{Belfiore}, F., {Maiolino}, R., {Maraston}, C., {et~al.} 2016, \mnras, 461,
	3111
	
	\bibitem[{{Binette} {et~al.}(1994){Binette}, {Magris}, {Stasi{\'n}ska}, \&
		{Bruzual}}]{Binette_1994_pAGB_ionization}
	{Binette}, L., {Magris}, C.~G., {Stasi{\'n}ska}, G., \& {Bruzual}, A.~G. 1994,
	\aap, 292, 13
	
	\bibitem[{{Brinchmann} {et~al.}(2004){Brinchmann}, {Charlot}, {White},
		{Tremonti}, {Kauffmann}, {Heckman}, \& {Brinkmann}}]{Brinchmann_2004_SDSS_Z}
	{Brinchmann}, J., {Charlot}, S., {White}, S.~D.~M., {et~al.} 2004, \mnras, 351,
	1151
	
	\bibitem[{{Calzetti} {et~al.}(1994){Calzetti}, {Kinney}, \&
		{Storchi-Bergmann}}]{Calzetti_1994_extinction}
	{Calzetti}, D., {Kinney}, A.~L., \& {Storchi-Bergmann}, T. 1994, \apj, 429, 582
	
	\bibitem[{{Castro} {et~al.}(2017){Castro}, {Dors}, {Cardaci}, \&
		{H{\"a}gele}}]{Castro_2017_AGN_Z}
	{Castro}, C.~S., {Dors}, O.~L., {Cardaci}, M.~V., \& {H{\"a}gele}, G.~F. 2017,
	\mnras, 467, 1507
	
	\bibitem[{{Charlot} \& {Longhetti}(2001)}]{Charlot_and_Longhetti_2001_SF_lines}
	{Charlot}, S., \& {Longhetti}, M. 2001, \mnras, 323, 887
	
	\bibitem[{{Cid Fernandes} {et~al.}(2011){Cid Fernandes}, {Stasi{\'n}ska},
		{Mateus}, \& {Vale Asari}}]{Cid_Fernandes_2011_WHAN}
	{Cid Fernandes}, R., {Stasi{\'n}ska}, G., {Mateus}, A., \& {Vale Asari}, N.
	2011, \mnras, 413, 1687
	
	\bibitem[{{Collin} \& {Zahn}(2008)}]{Collin_2008_Accretion_disc_SF}
	{Collin}, S., \& {Zahn}, J.-P. 2008, \aap, 477, 419
	
	\bibitem[{{Denicol{\'o}} {et~al.}(2002){Denicol{\'o}}, {Terlevich}, \&
		{Terlevich}}]{Denicolo_2002_N2}
	{Denicol{\'o}}, G., {Terlevich}, R., \& {Terlevich}, E. 2002, \mnras, 330, 69
	
	\bibitem[{{Dietrich} {et~al.}(2003){Dietrich}, {Hamann}, {Shields},
		{Constantin}, {Heidt}, {J{\"a}ger}, {Vestergaard}, \&
		{Wagner}}]{Dietrich_2003_BLR_Z}
	{Dietrich}, M., {Hamann}, F., {Shields}, J.~C., {et~al.} 2003, \apj, 589, 722
	
	\bibitem[{Dopita {et~al.}(2002)Dopita, Groves, Sutherland, Binette, \&
		Cecil}]{Dopita_2002_NLRs}
	Dopita, M.~A., Groves, B.~A., Sutherland, R.~S., Binette, L., \& Cecil, G.
	2002, The Astrophysical Journal, 572, 753.
	\newblock \url{http://iopscience.iop.org/0004-637X/572/2/753/fulltext/}
	
	\bibitem[{Dopita \& Sutherland(2003)}]{Dopita_Sutherland_2003_ISM_book}
	Dopita, M.~A., \& Sutherland, R.~S. 2003, {Astrophysics of the diffuse
		universe} (Berlin, New York: Springer).
	\newblock \url{http://adsabs.harvard.edu/abs/2003adu..book.....D}
	
	\bibitem[{{Dopita} {et~al.}(2006){Dopita}, {Fischera}, {Sutherland}, {Kewley},
		{Leitherer}, {Tuffs}, {Popescu}, {van Breugel}, \&
		{Groves}}]{Dopita_2006_SB_SED_III}
	{Dopita}, M.~A., {Fischera}, J., {Sutherland}, R.~S., {et~al.} 2006, \apjs,
	167, 177
	
	\bibitem[{Dopita {et~al.}(2014)Dopita, Scharw\"{a}chter, Shastri, Kewley,
		Davies, Sutherland, Kharb, Jose, Hampton, Jin, Banfield, Basurah, \&
		Fischer}]{Dopita_2014_probing_ENLR_I_S7}
	Dopita, M.~A., Scharw\"{a}chter, J., Shastri, P., {et~al.} 2014, Astronomy \&
	Astrophysics, 566, A41.
	\newblock \url{http://adsabs.harvard.edu/abs/2014A\%26A...566A..41D}
	
	\bibitem[{Dopita {et~al.}(2015)Dopita, Shastri, Davies, Kewley, Hampton,
		Scharw\"{a}chter, Sutherland, Kharb, Jose, Bhatt, Ramya, Jin, Banfield, Zaw,
		Juneau, James, \& Srivastava}]{Dopita_2015_probing_ENLR_II_S7}
	Dopita, M.~A., Shastri, P., Davies, R., {et~al.} 2015, \apjs, 217, 12.
	\newblock \url{http://adsabs.harvard.edu/abs/2015ApJS..217...12D}
	
	\bibitem[{{Dors} {et~al.}(2014){Dors}, {Cardaci}, {H{\"a}gele}, \&
		{Krabbe}}]{Dors_2014_AGN_Z}
	{Dors}, O.~L., {Cardaci}, M.~V., {H{\"a}gele}, G.~F., \& {Krabbe}, {\^A}.~C.
	2014, \mnras, 443, 1291
	
	\bibitem[{{Dors} {et~al.}(2015){Dors}, {Cardaci}, {H{\"a}gele}, {Rodrigues},
		{Grebel}, {Pilyugin}, {Freitas-Lemes}, \& {Krabbe}}]{Dors_2015_AGN_Z}
	{Dors}, O.~L., {Cardaci}, M.~V., {H{\"a}gele}, G.~F., {et~al.} 2015, \mnras,
	453, 4102
	
	\bibitem[{{Du} {et~al.}(2014){Du}, {Wang}, {Hu}, {Valls-Gabaud}, {Baldwin},
		{Ge}, \& {Xue}}]{Du_2014_NLR_BLR_Z}
	{Du}, P., {Wang}, J.-M., {Hu}, C., {et~al.} 2014, \mnras, 438, 2828
	
	\bibitem[{{Groves} {et~al.}(2004){Groves}, {Dopita}, \&
		{Sutherland}}]{Groves_2004_NLR_models_II}
	{Groves}, B.~A., {Dopita}, M.~A., \& {Sutherland}, R.~S. 2004, \apjs, 153, 75
	
	\bibitem[{{Groves} {et~al.}(2006){Groves}, {Heckman}, \&
		{Kauffmann}}]{Groves_2006_SDSS_AGN}
	{Groves}, B.~A., {Heckman}, T.~M., \& {Kauffmann}, G. 2006, \mnras, 371, 1559
	
	\bibitem[{{Hamann} \& {Ferland}(1999)}]{Hamann_and_Ferland_1999_QSO_Z_review}
	{Hamann}, F., \& {Ferland}, G. 1999, \araa, 37, 487
	
	\bibitem[{{Jenkins}(2009)}]{Jenkins_2009_depletions}
	{Jenkins}, E.~B. 2009, \apj, 700, 1299
	
	\bibitem[{{Jenkins}(2014)}]{Jenkins_2014_depletions}
	---. 2014, ArXiv e-prints, arXiv:1402.4765
	
	\bibitem[{{Juarez} {et~al.}(2009){Juarez}, {Maiolino}, {Mujica}, {Pedani},
		{Marinoni}, {Nagao}, {Marconi}, \& {Oliva}}]{Juarez_2009_QSO_Z}
	{Juarez}, Y., {Maiolino}, R., {Mujica}, R., {et~al.} 2009, \aap, 494, L25
	
	\bibitem[{{Kawasaki} {et~al.}(2017){Kawasaki}, {Nagao}, {Toba}, {Terao}, \&
		{Matsuoka}}]{Kawasaki_2017_SDSS_lowZ_AGN}
	{Kawasaki}, K., {Nagao}, T., {Toba}, Y., {Terao}, K., \& {Matsuoka}, K. 2017,
	\apj, 842, 44
	
	\bibitem[{{Kewley} \& {Dopita}(2002)}]{Kewley_and_Dopita_2002_Z}
	{Kewley}, L.~J., \& {Dopita}, M.~A. 2002, \apjs, 142, 35
	
	\bibitem[{{Kewley} {et~al.}(2001){Kewley}, {Dopita}, {Sutherland}, {Heisler},
		\& {Trevena}}]{Kewley2001}
	{Kewley}, L.~J., {Dopita}, M.~A., {Sutherland}, R.~S., {Heisler}, C.~A., \&
	{Trevena}, J. 2001, \apj, 556, 121
	
	\bibitem[{{Kewley} \& {Ellison}(2008)}]{Kewley_Ellison_2008_Z}
	{Kewley}, L.~J., \& {Ellison}, S.~L. 2008, \apj, 681, 1183
	
	\bibitem[{{Kewley} {et~al.}(2006){Kewley}, {Groves}, {Kauffmann}, \&
		{Heckman}}]{Kewley2006_AGN_hosts}
	{Kewley}, L.~J., {Groves}, B., {Kauffmann}, G., \& {Heckman}, T. 2006, \mnras,
	372, 961
	
	\bibitem[{{Lacerda} {et~al.}(2018){Lacerda}, {Cid Fernandes}, {Couto},
		{Stasi{\'n}ska}, {Garc{\'{\i}}a-Benito}, {Vale Asari}, {P{\'e}rez},
		{Gonz{\'a}lez Delgado}, {S{\'a}nchez}, \& {de
			Amorim}}]{Lacerda_2018_CALIFA_DIG}
	{Lacerda}, E.~A.~D., {Cid Fernandes}, R., {Couto}, G.~S., {et~al.} 2018,
	\mnras, 474, 3727
	
	\bibitem[{{Lequeux} {et~al.}(1979){Lequeux}, {Peimbert}, {Rayo}, {Serrano}, \&
		{Torres-Peimbert}}]{Lequeux_1979_MZ}
	{Lequeux}, J., {Peimbert}, M., {Rayo}, J.~F., {Serrano}, A., \&
	{Torres-Peimbert}, S. 1979, \aap, 80, 155
	
	\bibitem[{{Lilly} {et~al.}(2013){Lilly}, {Carollo}, {Pipino}, {Renzini}, \&
		{Peng}}]{Lilly_2013_bathtub}
	{Lilly}, S.~J., {Carollo}, C.~M., {Pipino}, A., {Renzini}, A., \& {Peng}, Y.
	2013, \apj, 772, 119
	
	\bibitem[{{Matsuoka} {et~al.}(2018){Matsuoka}, {Nagao}, {Marconi}, {Maiolino},
		{Mannucci}, {Cresci}, {Terao}, \& {Ikeda}}]{Matsuoka_2018_NLR_MZ}
	{Matsuoka}, K., {Nagao}, T., {Marconi}, A., {et~al.} 2018, \aap, 616, L4
	
	\bibitem[{{McGaugh}(1991)}]{McGaugh_1991_Z}
	{McGaugh}, S.~S. 1991, \apj, 380, 140
	
	\bibitem[{{Morisset} {et~al.}(2004){Morisset}, {Schaerer}, {Bouret}, \&
		{Martins}}]{Morisset_2004_ionizing_stellar_spectra}
	{Morisset}, C., {Schaerer}, D., {Bouret}, J.-C., \& {Martins}, F. 2004, \aap,
	415, 577
	
	\bibitem[{{Moustakas} {et~al.}(2011){Moustakas}, {Zaritsky}, {Brown}, {Cool},
		{Dey}, {Eisenstein}, {Gonzalez}, {Jannuzi}, {Jones}, {Kochanek}, {Murray}, \&
		{Wild}}]{Moustakas_2011_MZ_evolution}
	{Moustakas}, J., {Zaritsky}, D., {Brown}, M., {et~al.} 2011, ArXiv e-prints,
	arXiv:1112.3300
	
	\bibitem[{{Nagao} {et~al.}(2006{\natexlab{a}}){Nagao}, {Maiolino}, \&
		{Marconi}}]{Nagao_2006_AGN_NLR_Z}
	{Nagao}, T., {Maiolino}, R., \& {Marconi}, A. 2006{\natexlab{a}}, \aap, 447,
	863
	
	\bibitem[{{Nagao} {et~al.}(2006{\natexlab{b}}){Nagao}, {Maiolino}, \&
		{Marconi}}]{Nagao_2006_HII_Z}
	---. 2006{\natexlab{b}}, \aap, 459, 85
	
	\bibitem[{{Nagao} {et~al.}(2006{\natexlab{c}}){Nagao}, {Marconi}, \&
		{Maiolino}}]{Nagao_2006_SDSS_BLR_Z}
	{Nagao}, T., {Marconi}, A., \& {Maiolino}, R. 2006{\natexlab{c}}, \aap, 447,
	157
	
	\bibitem[{{Nayakshin} \& {Sunyaev}(2005)}]{Nayakshin_2005_SgrAstar_disk_SF}
	{Nayakshin}, S., \& {Sunyaev}, R. 2005, \mnras, 364, L23
	
	\bibitem[{{Nicholls} {et~al.}(2017){Nicholls}, {Sutherland}, {Dopita},
		{Kewley}, \& {Groves}}]{Nicholls_2017_abundance_scaling}
	{Nicholls}, D.~C., {Sutherland}, R.~S., {Dopita}, M.~A., {Kewley}, L.~J., \&
	{Groves}, B.~A. 2017, \mnras, 466, 4403
	
	\bibitem[{Osterbrock \& Ferland(2006)}]{Osterbrock_Ferland_2006_Gas_astro}
	Osterbrock, D., \& Ferland, G. 2006, Astrophysics of Gaseous Nebulae and Active
	Galactic Nuclei, 2nd edn. (University Science Books).
	\newblock \url{https://books.google.com.au/books?id=HgfrkDjBD98C}
	
	\bibitem[{{Pettini} \& {Pagel}(2004)}]{Pettini_and_Pagel_2004_Z_diagnostic}
	{Pettini}, M., \& {Pagel}, B.~E.~J. 2004, \mnras, 348, L59
	
	\bibitem[{{Pilyugin} {et~al.}(2018){Pilyugin}, {Grebel}, {Zinchenko},
		{Nefedyev}, {Shulga}, {Wei}, \& {Berczik}}]{Pilyugin_2018_DIG_contamination}
	{Pilyugin}, L.~S., {Grebel}, E.~K., {Zinchenko}, I.~A., {et~al.} 2018, \aap,
	613, A1
	
	\bibitem[{{Sanders} {et~al.}(2015){Sanders}, {Shapley}, {Kriek}, {Reddy},
		{Freeman}, {Coil}, {Siana}, {Mobasher}, {Shivaei}, {Price}, \& {de
			Groot}}]{Sanders_2015_MOSDEF_MZ}
	{Sanders}, R.~L., {Shapley}, A.~E., {Kriek}, M., {et~al.} 2015, \apj, 799, 138
	
	\bibitem[{{Storchi-Bergmann} {et~al.}(1998){Storchi-Bergmann}, {Schmitt},
		{Calzetti}, \& {Kinney}}]{Storchi-Bergmann_1998_AGN_Z}
	{Storchi-Bergmann}, T., {Schmitt}, H.~R., {Calzetti}, D., \& {Kinney}, A.~L.
	1998, \aj, 115, 909
	
	\bibitem[{{Sutherland} \&
		{Dopita}(2017)}]{Sutherland_Dopita_2017_shocks_MAPPINGSV}
	{Sutherland}, R.~S., \& {Dopita}, M.~A. 2017, \apjs, 229, 34
	
	\bibitem[{{Thomas} {et~al.}(2018){Thomas}, {Dopita}, {Kewley}, {Groves},
		{Sutherland}, {Hopkins}, \& {Blanc}}]{Thomas_2018_NebulaBayes}
	{Thomas}, A.~D., {Dopita}, M.~A., {Kewley}, L.~J., {et~al.} 2018, \apj, 856, 89
	
	\bibitem[{{Thomas} {et~al.}(2016){Thomas}, {Groves}, {Sutherland}, {Dopita},
		{Kewley}, \& {Jin}}]{Thomas_2016_oxaf}
	{Thomas}, A.~D., {Groves}, B.~A., {Sutherland}, R.~S., {et~al.} 2016, \apj,
	833, 266
	
	\bibitem[{{Thomas} {et~al.}(2017){Thomas}, {Dopita}, {Shastri}, {Davies},
		{Hampton}, {Kewley}, {Banfield}, {Groves}, {James}, {Jin}, {Juneau}, {Kharb},
		{Sairam}, {Scharw{\"a}chter}, {Shalima}, {Sundar}, {Sutherland}, \&
		{Zaw}}]{Thomas_2017_S7_DR2}
	{Thomas}, A.~D., {Dopita}, M.~A., {Shastri}, P., {et~al.} 2017, \apjs, 232, 11
	
	\bibitem[{{Tremonti} {et~al.}(2004){Tremonti}, {Heckman}, {Kauffmann},
		{Brinchmann}, {Charlot}, {White}, {Seibert}, {Peng}, {Schlegel}, {Uomoto},
		{Fukugita}, \& {Brinkmann}}]{Tremonti_2004_MZ}
	{Tremonti}, C.~A., {Heckman}, T.~M., {Kauffmann}, G., {et~al.} 2004, \apj, 613,
	898
	
	\bibitem[{{Veilleux} \& {Osterbrock}(1987)}]{Veilleux_1987_LineRatios}
	{Veilleux}, S., \& {Osterbrock}, D.~E. 1987, \apjs, 63, 295
	
	\bibitem[{{Vogt} {et~al.}(2013){Vogt}, {Dopita}, \& {Kewley}}]{Vogt_2013_HCG_I}
	{Vogt}, F.~P.~A., {Dopita}, M.~A., \& {Kewley}, L.~J. 2013, \apj, 768, 151
	
	\bibitem[{{Wuyts} {et~al.}(2016){Wuyts}, {Wisnioski}, {Fossati}, {F{\"o}rster
			Schreiber}, {Genzel}, {Davies}, {Mendel}, {Naab}, {R{\"o}ttgers}, {Wilman},
		{Wuyts}, {Bandara}, {Beifiori}, {Belli}, {Bender}, {Brammer}, {Burkert},
		{Chan}, {Galametz}, {Kulkarni}, {Lang}, {Lutz}, {Momcheva}, {Nelson},
		{Rosario}, {Saglia}, {Seitz}, {Tacconi}, {Tadaki}, {{\"U}bler}, \& {van
			Dokkum}}]{Wuyts_2016_highz_Z}
	{Wuyts}, E., {Wisnioski}, E., {Fossati}, M., {et~al.} 2016, \apj, 827, 74
	
	\bibitem[{{York} {et~al.}(2000){York}, {Adelman}, {Anderson}, {Anderson},
		{Annis}, {Bahcall}, {Bakken}, {Barkhouser}, {Bastian}, {Berman}, {Boroski},
		{Bracker}, {Briegel}, {Briggs}, {Brinkmann}, {Brunner}, {Burles}, {Carey},
		{Carr}, {Castander}, {Chen}, {Colestock}, {Connolly}, {Crocker}, {Csabai},
		{Czarapata}, {Davis}, {Doi}, {Dombeck}, {Eisenstein}, {Ellman}, {Elms},
		{Evans}, {Fan}, {Federwitz}, {Fiscelli}, {Friedman}, {Frieman}, {Fukugita},
		{Gillespie}, {Gunn}, {Gurbani}, {de Haas}, {Haldeman}, {Harris}, {Hayes},
		{Heckman}, {Hennessy}, {Hindsley}, {Holm}, {Holmgren}, {Huang}, {Hull},
		{Husby}, {Ichikawa}, {Ichikawa}, {Ivezi{\'c}}, {Kent}, {Kim}, {Kinney},
		{Klaene}, {Kleinman}, {Kleinman}, {Knapp}, {Korienek}, {Kron}, {Kunszt},
		{Lamb}, {Lee}, {Leger}, {Limmongkol}, {Lindenmeyer}, {Long}, {Loomis},
		{Loveday}, {Lucinio}, {Lupton}, {MacKinnon}, {Mannery}, {Mantsch}, {Margon},
		{McGehee}, {McKay}, {Meiksin}, {Merelli}, {Monet}, {Munn}, {Narayanan},
		{Nash}, {Neilsen}, {Neswold}, {Newberg}, {Nichol}, {Nicinski}, {Nonino},
		{Okada}, {Okamura}, {Ostriker}, {Owen}, {Pauls}, {Peoples}, {Peterson},
		{Petravick}, {Pier}, {Pope}, {Pordes}, {Prosapio}, {Rechenmacher}, {Quinn},
		{Richards}, {Richmond}, {Rivetta}, {Rockosi}, {Ruthmansdorfer}, {Sandford},
		{Schlegel}, {Schneider}, {Sekiguchi}, {Sergey}, {Shimasaku}, {Siegmund},
		{Smee}, {Smith}, {Snedden}, {Stone}, {Stoughton}, {Strauss}, {Stubbs},
		{SubbaRao}, {Szalay}, {Szapudi}, {Szokoly}, {Thakar}, {Tremonti}, {Tucker},
		{Uomoto}, {Vanden Berk}, {Vogeley}, {Waddell}, {Wang}, {Watanabe},
		{Weinberg}, {Yanny}, {Yasuda}, \& {SDSS Collaboration}}]{York_2000_SDSS}
	{York}, D.~G., {Adelman}, J., {Anderson}, Jr., J.~E., {et~al.} 2000, \aj, 120,
	1579
	
	\bibitem[{{Zahid} {et~al.}(2014{\natexlab{a}}){Zahid}, {Dima}, {Kudritzki},
		{Kewley}, {Geller}, {Hwang}, {Silverman}, \& {Kashino}}]{Zahid_2014_MZ}
	{Zahid}, H.~J., {Dima}, G.~I., {Kudritzki}, R.-P., {et~al.} 2014{\natexlab{a}},
	\apj, 791, 130
	
	\bibitem[{{Zahid} {et~al.}(2014{\natexlab{b}}){Zahid}, {Kashino}, {Silverman},
		{Kewley}, {Daddi}, {Renzini}, {Rodighiero}, {Nagao}, {Arimoto}, {Sanders},
		{Kartaltepe}, {Lilly}, {Maier}, {Geller}, {Capak}, {Carollo}, {Chu},
		{Hasinger}, {Ilbert}, {Kajisawa}, {Koekemoer}, {Kovacs}, {Le F{\`e}vre},
		{Masters}, {McCracken}, {Onodera}, {Scoville}, {Strazzullo}, {Sugiyama},
		{Taniguchi}, \& {COSMOS Team}}]{Zahid_2014_MZ_II}
	{Zahid}, H.~J., {Kashino}, D., {Silverman}, J.~D., {et~al.} 2014{\natexlab{b}},
	\apj, 792, 75
	
	\bibitem[{{Zaritsky} {et~al.}(1994){Zaritsky}, {Kennicutt}, \&
		{Huchra}}]{Zaritsky_1994_Luminosity_Z}
	{Zaritsky}, D., {Kennicutt}, Jr., R.~C., \& {Huchra}, J.~P. 1994, \apj, 420, 87
	
	\bibitem[{{Zhang} {et~al.}(2017){Zhang}, {Yan}, {Bundy}, {Bershady}, {Haffner},
		{Walterbos}, {Maiolino}, {Tremonti}, {Thomas}, {Drory}, {Jones}, {Belfiore},
		{S{\'a}nchez}, {Diamond-Stanic}, {Bizyaev}, {Nitschelm}, {Andrews},
		{Brinkmann}, {Brownstein}, {Cheung}, {Li}, {Law}, {Roman Lopes}, {Oravetz},
		{Pan}, {Storchi Bergmann}, \& {Simmons}}]{Zhang_2017_MaNGA_DIG}
	{Zhang}, K., {Yan}, R., {Bundy}, K., {et~al.} 2017, \mnras, 466, 3217
	
\end{thebibliography}
\end{document}